\pdfoutput=1
\documentclass[smallabstract,smallcaptions]{dccpaper}
\ifx\HCode\UnDef\else\hypersetup{tex4ht}\fi
\usepackage[english]{babel}
\usepackage[utf8]{inputenc}

\usepackage{array}

\usepackage{amsmath,amsfonts,amssymb}
\usepackage[natbib,cref,theorems,pdf]{nikis}

\usepackage{color}
\usepackage{url}
\usepackage{booktabs}
\usepackage{enumerate}

\usepackage{ifthen}
\newboolean{isJournal} 
\setboolean{isJournal}{true}
\ifthenelse{\boolean{isJournal}}{%

	\newcommand{\JO}[2][]{#2}
	\newcommand{\Jitemize}[1]{\sitemize{#1}}

}{%
	\usepackage{etoolbox}

	\newcommand{\JO}[2][]{#1}
	\newcommand{\Jitemize}[1]{\sitemize*{#1}}
}
\let\OLDthebibliography\thebibliography{}
\renewcommand\thebibliography[1]{%
  \OLDthebibliography{#1}
  \setlength{\parskip}{0pt}
  \setlength{\itemsep}{0pt plus 0.3ex}
}

\usepackage{xcolor} %

 \definecolor{solarizedBase03}{HTML}{002B36}
 \definecolor{solarizedBase02}{HTML}{073642}
 \definecolor{solarizedBase01}{HTML}{586e75}
 \definecolor{solarizedBase00}{HTML}{657b83}
 \definecolor{solarizedBase0}{HTML}{839496}
 \definecolor{solarizedBase1}{HTML}{93a1a1}
 \definecolor{solarizedBase2}{HTML}{EEE8D5}
 \definecolor{solarizedBase3}{HTML}{FDF6E3}
 \definecolor{solarizedYellow}{HTML}{B58900}
 \definecolor{solarizedOrange}{HTML}{CB4B16}
 \definecolor{solarizedRed}{HTML}{DC322F}
 \definecolor{solarizedMagenta}{HTML}{D33682}
 \definecolor{solarizedViolet}{HTML}{6C71C4}
 \definecolor{solarizedBlue}{HTML}{268BD2}
 \definecolor{solarizedCyan}{HTML}{2AA198}
 \definecolor{solarizedGreen}{HTML}{859900}

 \colorlet{takeLeaf}{black} %
\colorlet{LZNode}{black}%

\usepackage{algorithm}
\usepackage{algorithmic}

\newcommand{\BREAK}{\STATE \textbf{break}}
\newcommand{\CONTINUE}{\STATE \textbf{continue}}
\newcommand{\INCR}[1]{\STATE \textbf{incr} #1}

\newlength{\figurewidth}
\newlength{\smallfigurewidth}

\newcommand{\numWitness}{\ensuremath{z_\mathup{W}}}
\newcommand{\numRefs}{\ensuremath{z_\mathup{R}}}

\newcommand{\exampleString}{\JO[{{\tt aabaababa\$}}]{{\tt aaababaaabaaba\$}}}

\newcommand{\rootnode}     {\instancename{root}}

\newcommand{\ISA}          {\ensuremath{\instancename{SA}^{-1}}}

\newcommand{\SA}           {\instancename{SA}}

\newcommand{\ST}           {\instancename{ST}}

\newcommand{\smallestleaf}[1][]{\UnaryOperator[#1]{\textsl{smallest\_leaf}}}
\newcommand{\levelanc}    [1][]{\UnaryOperator[#1]{\textsl{level\_anc}}}
\newcommand{\strdepth}    [1][]{\UnaryOperator[#1]{\textsl{str\_depth}}}
\newcommand{\depth}       [1][]{\UnaryOperator[#1]{\textsl{depth}}}
\newcommand{\nextleaf}    [1][]{\UnaryOperator[#1]{\textsl{next\_leaf}}}

\newcommand{\child}       [1][]{\UnaryOperator[#1]{\textsl{child}}}
\newcommand{\head}        [1][]{\UnaryOperator[#1]{\textsl{head}}}
\newcommand{\descendant}  [1][]{\UnaryOperator[#1]{\textsl{descendant}}}
\newcommand{\popcount}    [1][]{\UnaryOperator[#1]{\textsl{popcount}}}
\newcommand{\lab}         [1][]{\UnaryOperator[#1]{\textsl{label}}}
\newcommand{\parent}      [1][]{\UnaryOperator[#1]{\textsl{parent}}}
\newcommand{\childrank}   [1][]{\UnaryOperator[#1]{\textsl{child\_rank}}}
\newcommand{\leafselect}  [1][]{\UnaryOperator[#1]{\textsl{leaf\_select}}}
\newcommand{\leafrank}  [1][]{\UnaryOperator[#1]{\textsl{leaf\_rank}}}
\newcommand{\lmostleaf}   [1][]{\UnaryOperator[#1]{\textsl{lmost\_leaf}}}
\newcommand{\rmostleaf}   [1][]{\UnaryOperator[#1]{\textsl{rmost\_leaf}}}

\newcommand{\clear}      {\textsl{clear}}
\newcommand{\addRank}      {\textsl{add\_rank\_support}}

\newcommand{\Entr}[1]      {\UnaryOperator[#1]{\textsl{H}_0}}

\usepackage{marvosym}

\usepackage{enumitem}

\newlist{passes}{enumerate}{10}
\setlist[passes]{label*=(\alph*)}

\crefname{passesi}{pass}{passes}
\Crefname{passesi}{Pass}{Passes}

\newcommand{\bv}[1]{\ensuremath{B_{\mathup{#1}}}}
\newcommand{\maru}{\ensuremath{\bigcirc}}
\newcommand{\Maru}{}

\begin{document}

\title%
{\large%
	\textbf{Lempel-Ziv Computation In Compressed Space (${\text{LZ-CICS}}^\text{\Football}$)}
}

\author{%
	Dominik K\"{o}ppl$^1$ and Kunihiko Sadakane$^2$
	\\[0.5em]
	$^1$Department of Computer Science, TU Dortmund, Germany
	\\
\hspace{-1em} $^2$Graduate School of Information Science and Technology, University of Tokyo, Japan \\
}

\maketitle
\thispagestyle{empty}

\begin{abstract}
	We show that both the Lempel-Ziv-77 and the Lempel-Ziv-78 factorization of a text of length~$n$ on an integer alphabet of size~$\sigma$ 
	can be computed in $\Oh{n \lg \lg \sigma}$ time (linear time if we allow randomization)
using $\Oh{n \lg \sigma}$ bits of working space.
Given that a compressed representation of the suffix tree is loaded into RAM, we can compute both factorizations in \Oh{n} time using
$z \lg n + \Oh{n}$ bits of space, where $z$ is the number of factors.
\end{abstract}

\Section{Introduction}
The Lempel-Ziv-77~(LZ77)~\cite{Ziv1977uas} and the Lempel-Ziv-78~(LZ78)~\cite{Ziv1978Coi} factorization 
divide a text into factors that capture repetitions in the text.
Although both factorizations are found in major text processing tools like compressors or full text indices,
computing any of the two factorizations is a bottleneck in terms of space and time.
In practice, compressors still use a sliding window or discarding techniques to avoid high resource consumption.
Hence, one might ask whether it is possible to lower the space bound in the light of recent approaches in the field of succinct data structures,
while still allowing linear running time.

In this article, we show that the LZ77 and the LZ78 factorization of a text of length~$n$ on an integer alphabet of size~$\sigma$ can be computed
\JO[%
	with $\Oh{n \lg \sigma}$ bits of working space in either $\Oh{n}$ randomized or $\Oh{n \lg \lg \sigma}$ deterministic time.]{
\begin{itemize}
	\item with $\Oh{n \lg \sigma}$ bits of working space in either $\Oh{n}$ randomized or $\Oh{n \lg \lg \sigma}$ deterministic time, and
	\item with $z\lg n + \Oh{n}$ additional bits of working space in linear time, given that we have access to the compressed suffix tree of the text.
\end{itemize}
}%

\Section{Related Work}

We are aware of the following results for LZ77:
The currently most space efficient algorithm is due to \citet{kosoblovLZ}
whose algorithm runs in $\Oh{n(\lg \sigma + \lg \lg n)}$ time and uses only $\varepsilon n$ bits of working space, 
provided that we have read-access to the text.
A trade-off algorithm is given by \citet{kempaLightLZ}, using $\Oh{n/d}$ words of working space and $\Oh{dn}$ time.
By setting $d \gets \log_{\sigma} n$ we get $\Oh{n \lg \sigma}$ bits of working space and $\Oh{n \log_{\sigma} n}$ time.
The algorithm of \citet{djamalLZ} derives its dominant terms in space and time from the same data structure~\cite{djamalCSTOld} as we do;
the LZ77 factorization algorithms of both papers work with the same space and time bounds.
Their algorithm uses only the Burrows-Wheeler transform~(BWT)~\cite{bwt} construction algorithm from~\cite{djamalCSTOld}.
By exchanging it with an improved version~\cite{djamalCST}, 
we expect that their algorithm will run in deterministic linear time.

Since LZ78 factors are naturally represented in a trie, the so-called \intWort{LZ trie},
improving LZ78 computation can be done, among others, by using sophisticated trie implementations~\cite{fischer15alphabet,Jansson2015LDT},
or by superimposing the suffix tree with the suffix trie~\cite{Nakashima2015CLT,lzciss}.
We follow the latter approach.
There, both \citet{Nakashima2015CLT} and \citet{lzciss} presented a linear time algorithm, using 
$\Oh{n \lg n}$ and $(1+\varepsilon) n \lg n + \Oh{n}$ bits of space, respectively.

Based on the data structures of the compressed suffix tree, we derive our techniques 
from an approach~\cite{lzciss} using the suffix tree topology with succinct representations of the suffix array, its inverse and the longest common prefix array.
For both factorization variants, 
\citet{lzciss} store the inverse suffix array and parts of the enhanced suffix array in $(1+\varepsilon)n \lg n + \Oh{n}$ bits of space
such that
they can access leaves of the suffix tree in text order, and can compute the string depth of internal nodes, both in constant time. 
Unlike the here presented approach, their algorithms overwrite the working space multiple times, and use a complicated counting for the LZ78 trie nodes.

\Section{Preliminaries}
Our computational model is the word RAM model with word size $\Om{\lg n}$ for some natural number~$n$.
Accessing a word costs $\Oh{1}$ time.
We assume that the function \popcount[w], counting the set bits in a word $w$, can be computed in constant time.
Otherwise, we build a lookup-table~\cite{munro96tables} supporting \popcount{} with two rank queries in constant time.
The lookup-table fits into our working space.

Let $\Sigma$ denote an integer alphabet of size $\sigma = \abs{\Sigma} \le n$.
We call an element $T \in \Sigma^*$ a \intWort{string} or \intWort{text}.
Its length is denoted by $\abs{T}$.
The empty string is $\epsilon$ with $\abs{\epsilon}=0$.
We access the $j$-th character of~$T$ with $T[j]$ for $1 \le j \le \abs{T}$.
Given $x,y,z \in \Sigma^*$ with $T = xyz$, 
we call $x$, $y$, and $z$ a \intWort{prefix}, a \intWort{substring}, and a \intWort{suffix} of $T$, respectively.
In particular, the suffix starting at position~$j$ of~$T$ is called the \intWort{$j$-th suffix} of~$T$.

We call strings on the binary alphabet $\menge{0,1}$ \intWort{bit vectors}.
For a bit vector~$\bv{}$, we are interested in answering the following queries for $c \in \menge{0,1} \cup \menge{0,1}^2$:
\begin{itemize}
\JO[\crampedItems{}]{}
	\item $\bv{}.\rank[c](j)$ counts the number of \bsq{c}s in $\bv{}[1,j]$, and 
	\item $\bv{}.\select[c](j)$ gives the position of the $j$-th \bsq{c} in $\bv{}$.
\end{itemize}
We can answer both types of queries due to a result of \citet{rrrBV}:
There is a data structure taking $\oh{\abs{\bv{}}}$ extra bits of space 
to answer $\rank{}$ and $\select{}$ queries in constant time.
It can be constructed in time linear to $\abs{\bv{}}$.
We say that a bit vector has a \intWort{rank-support} and a \intWort{select-support} if it provides constant time access to 
$\rank$ and $\select$, respectively.

\JO{%
The zero-order entropy~$\Entr{n,z}$ of a bit vector of length $n$ storing $z$ ones
is defined by
$\Entr{n,z} n =  z \lg(n/z) + (n-z) \lg\left( n/(n-z)\right)$.
Such a bit vector can be compressed such that it consumes $\Entr{n,z}n + \oh{n}$ bits,
supporting access and rank in constant time~(e.g.,~\cite{rrrBV}).
}%

In the rest of this paper, we take a read-only text $T$ of length $n$,
which is subject to the LZ77 or the LZ78 factorization.
Let $T[n]$ be a special character appearing nowhere else in~$T$,
so that no suffix of~$T$ is a prefix of another suffix of~$T$. 
Without loss of generality, 
we assume that $\Sigma$ is the \emph{effective} alphabet of $T$, i.e., each character of $\Sigma$ appears in $T$ at least once.
Otherwise, we can reduce the alphabet to the effective alphabet by sorting the characters with a linear time integer sorting algorithm using 
$n \lg \sigma + \Oh{1}$~working space~\cite{RadixSortNoExtra}, and an array with $\sigma \lg \sigma$ bits to reconstruct the former alphabet.

\SubSection{Lempel-Ziv Factorization}
A \intWort{factorization} of $T$ with size~$z$ partitions $T$ into $z$~substrings $T=f_1 \cdots f_z$.
These substrings are called \intWort{factors}. In particular, we have:

\begin{definition}\label{defLZ77}
A factorization $f_1\cdots f_z = T$ is called the \intWort{LZ77 factorization} of $T$ iff 
$f_x = \argmax_{S \in S_j(T) \cup \Sigma} \abs{S}$ 
for all $1 \le x \le z$ with $j = \abs{f_1\cdots f_{x-1}}+1$,
where $S_j(T)$ denotes the set of substrings of $T$ that start strictly before~$j$ (for $1 \le j \le \abs{T}$).
\end{definition}

\JO[]{%
Different to the LZ77 factorization, the classic-LZ77 factorization adds an additional character to the referencing factors:
\begin{definition}
A factorization $f_1\cdots f_z = T$ is called the \intWort{classic-LZ77 factorization} of $T$ iff 
$f_x$ is the shortest prefix of $f_x \cdots f_z$ that occurs exactly once in $f_1 \cdots f_x$.
\end{definition}
}

\begin{definition}
A factorization $f_1\cdots f_z = T$ is called the \intWort{LZ78 factorization} of $T$ iff 
$f_x=f'_x\cdot c$ with $f'_x = \argmax_{S \in \menge{f_y : y < x} \cup \menge{\epsilon} } \abs{S}$ and $c\in\Sigma$
for all $1 \le x \le z$.
\end{definition}

\SubSection{Suffix Tree}

\begin{figure}[t]

	\JO[ {%
		\begin{tabular}{@{}*{2}{m{0.4\textwidth}}}
		\includegraphics[scale=1.0]{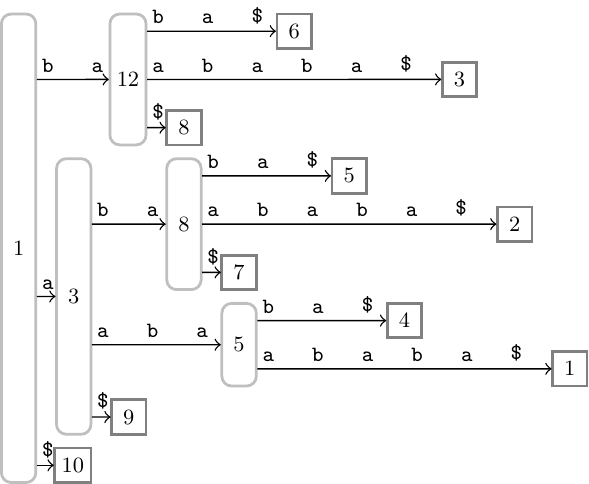}
&
		{\small
	\begin{tabular}{rcccccccccc}
		$i$      & {\tt 1}  & {\tt 2} & {\tt 3} & {\tt 4} & {\tt 5}  & {\tt 6}  & {\tt 7} & {\tt 8} & {\tt 9} & {\tt 10} \\
		\toprule
		$T$      & {\tt a}  & {\tt a} & {\tt b} & {\tt a} & {\tt a}  & {\tt b}  & {\tt a} & {\tt b} & {\tt a} & {\tt \$}\\
		\midrule
		$\SA$    & {\tt 10} & {\tt 9} & {\tt 1} & {\tt 4} & {\tt 7}  & {\tt 2}  & {\tt 5} & {\tt 8} & {\tt 3} & {\tt 6} \\
		\midrule
$\SA^{-{\tt 1}}$       & {\tt 3}  & {\tt 6} & {\tt 9} & {\tt 4} & {\tt 7}  & {\tt 10} & {\tt 5} & {\tt 8} & {\tt 2} & {\tt 1} \\
		\midrule
		$\psi$ &  	{\tt 3} & {\tt 1} & {\tt 6} & {\tt 7} & {\tt 8} & {\tt 9} & {\tt 10} & {\tt 2} & {\tt 4} & {\tt 5} \\
   \bottomrule
	\end{tabular}

	\begin{tabular}{rl}
		\\
		BP & {\tt (()(()(()())(()()()))(()()()))} \\
		leaves & {\tt 010010010100010101000010101000} \\
	\end{tabular}
}
\end{tabular}
\caption{The suffix tree of $T = \exampleString$. Internal nodes are labeled by their pre-order numbers, 
		leaves by the text position where their respective suffix starts.
		The number of letters on an edge~$e$ is $c(e)$. 
		Leaves are given in the BP representation by \bsq{{\tt ()}}. 
		By creating rank- and select-supports on \bsq{{\tt ()}} and \bsq{{\tt (}}, we can access internal nodes and leaves separately.
	}
} ]{\centering%
	\begin{tabular}{@{}*{2}{m{0.45\linewidth}}}
		\includegraphics[scale=1.0]{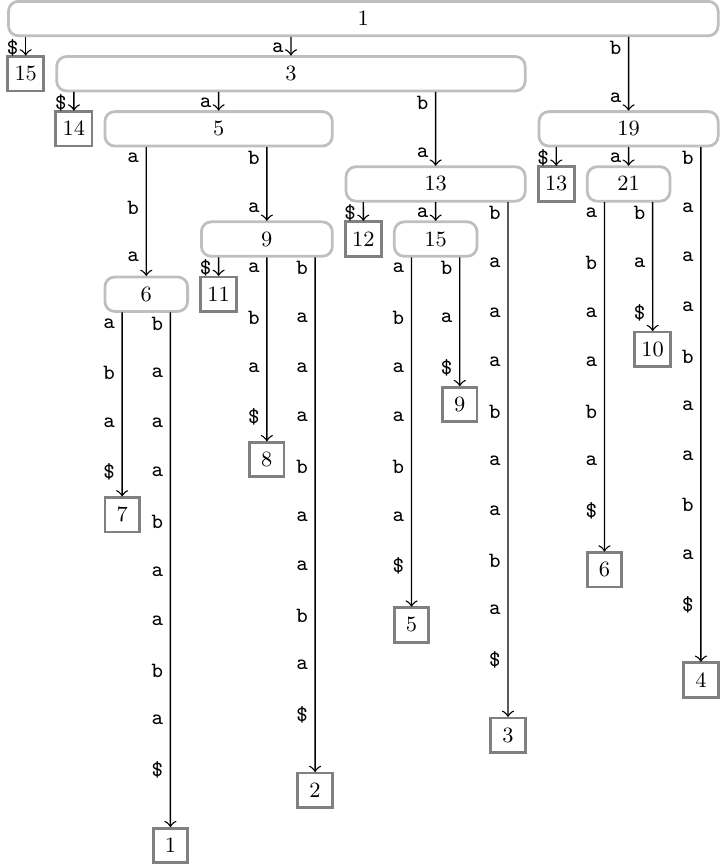}
		&\hspace{1em}
		{\fontsize{9}{10}\selectfont{}
	\tabcolsep=0.1em
	\begin{tabular}{r*{15}{c}}
		$i$        & {\tt 1}  & {\tt 2}  & {\tt 3}  & {\tt 4}  & {\tt 5}  & {\tt 6}  & {\tt 7}  & {\tt 8}  & {\tt 9}  & {\tt 10} & {\tt 11} & {\tt 12} & {\tt 13} & {\tt 14} & {\tt 15}
		\\\toprule
		$T$        & {\tt a}  & {\tt a}  & {\tt a}  & {\tt b}  & {\tt a}  & {\tt b}  & {\tt a}  & {\tt a}  & {\tt a}  & {\tt b}  & {\tt a}  & {\tt a}  & {\tt b}  & {\tt a}  & {\tt \$}
		\\\midrule
		$\SA$      & {\tt 15} & {\tt 14} & {\tt 7}  & {\tt 1}  & {\tt 11} & {\tt 8}  & {\tt 2}  & {\tt 12} & {\tt 5}  & {\tt 9}  & {\tt 3}  & {\tt 13} & {\tt 6}  & {\tt 10} & {\tt 4}
		\\\midrule
		$\SA^{-1}$ & {\tt 4}  & {\tt 7}  & {\tt 11} & {\tt 15} & {\tt 9}  & {\tt 13} & {\tt 3}  & {\tt 6}  & {\tt 10} & {\tt 14} & {\tt 5}  & {\tt 8}  & {\tt 12} & {\tt 2}  & {\tt 1}
		\\\midrule
		$\psi$     & {\tt 4}  & {\tt 1}  & {\tt 6}  & {\tt 7}  & {\tt 8}  & {\tt 10} & {\tt 11} & {\tt 12} & {\tt 13} & {\tt 14} & {\tt 15} & {\tt 2}  & {\tt 3}  & {\tt 5}  & {\tt 9}
  \\ \bottomrule
	\end{tabular}
}{\fontsize{7}{10}\selectfont{}
	\begin{tabular}{rl}
		\\
		BP &     {\tt (()(()((()())(()()()))(()(()())()))(()(()())()))} \\
		leaves &     {\tt 010010001010001010100001001010010000100101001000}
	\end{tabular}
}%
\end{tabular}
	\caption{The suffix tree of $T = \exampleString$. Internal nodes are labeled by their pre-order numbers, 
		leaves by the text position where their respective suffix starts.
		The number of letters on an edge~$e$ is $c(e)$. 
		\JO[Leaves are given in the BP representation by \bsq{{\tt ()}}.]{Nodes and leaves are represented by \bsq{{\tt (}} and \bsq{{\tt ()}}, respectively, in the BP representation.}
		By creating rank- and select-supports on \bsq{{\tt ()}} and \bsq{{\tt (}}, we can access internal nodes and leaves separately.
	}
}
\label{figST}
\end{figure}

The \intWort{suffix trie} of $T$ is the trie of all suffixes of $T$.
The \intWort{suffix tree~(ST)} of $T$, denoted by \ST{}, is the tree obtained by compacting the suffix trie of $T$.
We denote the root node of \ST{} by \rootnode{}.
Our approach uses a compressed representation of \ST, consisting of
\begin{itemize}
\JO[\crampedItems{}]{}
	\item the $\psi$-array~\cite{phiarray} with $\SA[i] = \SA[\psi(i)] - 1$ for $1 \le i \le n$ with $\SA[i] \not= n$ (and $\psi(i) = \SA^{-1}[1]$ for $\SA[i]=n$), and
	\item a $4n + \oh{n}$-bit balanced parenthesis representation (BP)\JO{~\cite{jacobson89space}} of the tree topology~\cite{cstsada}, 
		equipped with the minmax tree~\cite{statdynminmax} for navigation.
\end{itemize}
By employing the algorithm of~\citet{djamalCSTOld} on the text, 
we can build the compressed suffix tree consuming $\Oh{n \lg \sigma}$ bits of space in either $\Oh{n}$ randomized time or 
$\Oh{n \lg \lg \sigma}$ deterministic time.

Due to the BP representation, each node of the suffix tree is uniquely identified by its pre-order number.
A rank- and a select-support on the BP representation enable us to address a node by its pre-order number in constant time.
If the context is clear, we implicitly convert an \ST{} node to its pre-order number, and vice versa.

Each leaf is labeled \emph{conceptually} by the text position where its corresponding suffix starts~(see \Cref{figST}).
We write $\lab[\ell]$ for the label of a leaf~$\ell$.
Reading the leaf labels in depth first order returns the suffix array,
which we denote by \SA{}.
We do neither store \SA{} nor the leaf labels.

For descriptive purposes, we define the conceptional function $c(e)$ returning, for each edge~$e$, the length of~$e$.

We use the following methods on the ST topology that are well known to be computable in constant time after suitable preprocessing (see~\cite{statdynminmax}):
\Jitemize{%
\item  $\parent[v]$ selects the parent of the node~$v$,
\item  $\depth[v]$ returns the depth of the node~$v$,
\item  $\levelanc[\ell,d]$ selects the ancestor of the leaf~$\ell$ at depth~$d$,
\JO{\item  $\lmostleaf(v)$ and $\rmostleaf(v)$ selects the leftmost and rightmost leaf of node~$v$, respectively,}
\item  $\leafselect[i]$ selects the $i$-th leaf,
\item  $\leafrank[\ell]$ returns the number of preceding leaves of the leaf~$\ell$,
\item  $\childrank[v]$ returns the number of preceding siblings of the node~$v$, and
\item  $v.\child[i]$ selects the $i$-th child of the node~$v$.
}

Besides those basic tools, we need the following supplementary functions whose implementation details follow their descriptions:

\Jitemize{%
	\item  $\head[\ell]$ retrieves the first character of the suffix whose starting position coincides with the label of the leaf~$\ell$.
		Since $\Sigma$ is the effective alphabet of~$T$, each character of~$\Sigma$ occurs in~$T$.
		So \rootnode{} has $\sigma$~children, each corresponding to a different character.
		Besides, the order of \rootnode{}'s children and of the characters of~$\Sigma$ is the same.
		Hence, $\childrank[{\levelanc[\ell,{\depth[\rootnode]} +1]} ] = \head[\ell]$ holds, and the left hand side can be computed in constant time.

	\item  $\smallestleaf$ selects the leaf with the label $1$.
	  By a linear scan over the $\psi$-array, we can find the value $\alpha := \ISA[1]$, so that $\psi^{k}[\alpha] = \ISA[k+1]$ for $0 \le k \le n-1$.
We store $\alpha$ to answer $\smallestleaf$ by $\leafselect[\alpha]$.

\item  $\nextleaf[\ell]$ selects the leaf labeled with $\lab[\ell]+1$.
	We can compute it in constant time, since $\nextleaf[\ell] = \leafselect[{\psi[\leafrank[\ell] ]} ]$.

\item  $\strdepth[v]$ returns the string depth of an \emph{internal} node.
		We use the $\psi$-array and the $\head$-function to compute $\strdepth[v]$ in time proportional to the string depth\JO{~(see~\Cref{algoStrdepth})}.
		Therefore, we take two different children of $v$ (they exist since $v$ is an internal node), 
		and choose an arbitrary leaf in the subtree of each child.
		So we have two leaves representing two suffixes whose longest common prefix 
		is the string read from the edge labels on the path from the root to the lowest common ancestor~(LCA) of both leaves.
		Our task is to compute the length of this prefix.
		To this end, we match the first characters of both suffixes by the $\head{}$-function.
		If they match, we use $\psi$ to move to the next pair of suffixes, and apply the $\head{}$-function again.
		Informally, applying $\psi$ strips the first character of both suffixes (like taking a suffix link).
		On a mismatch, we find the first pair of characters that does not belong to the path from the root to $v$ 
		(reading the labels of the edges along this path).
		We return the number of matched characters as the string depth.
}%

\JO{%
\begin{algorithm}[tb]
\begin{algorithmic}[1]
	\REQUIRE suffix tree node $v$ 
	\IF {$v$ is an internal node}
	\STATE $\ell \gets \lmostleaf[{v.\child[1]} ]$
	\STATE $\ell' \gets \lmostleaf[{v.\child[2]} ]$ \COMMENT{$v.\child[2]$ exists since $v$ is an internal node}
	\STATE $m \gets 0$
	\WHILE{$\head[\ell] = \head[\ell']$} 
	\STATE $\ell \gets \nextleaf[\ell]$
	\STATE $\ell' \gets \nextleaf[\ell']$
	\INCR $m$
	  \ENDWHILE
	\RETURN $m$
	\ELSIF[works only if {\lab[v]} is available] {$v$ is a leaf}  
	\RETURN $n + 1 - \lab[v]$ 
\ENDIF
\end{algorithmic}
\caption{Implementation of \strdepth{}}
\label{algoStrdepth}
\end{algorithm}
}%

\Section{Common Settings}

We identify factors by text positions, i.e., we call a text position $j$ the \intWort{factor position} of $f_x$ ($1\le x \le z$) iff 
the factor $f_x$ starts at position $j$.
A factor $f_x$ may refer to either (LZ77) a previous text position $j$ (called $f_x$'s \intWort{referred position}), or (LZ78) to a previous factor $f_y$ (called $f_x$'s \intWort{referred factor}---in this case $y$ is also called the \intWort{referred index} of $f_x$).
If there is no suitable reference found for a given factor $f_x$ with factor position $j$, then $f_x$ consists of just the single letter $T[j]$. 
We call such a factor a \intWort{fresh factor}.
The other factors are called \intWort{referencing factors}.
Let \numRefs{} denote the number of referencing factors.
An example is given in~\Cref{figParsing}. \JO[A more sophisticated example can be found in the full version of the paper~\cite{lzcicsArXiv}.]{}

\begin{figure}[t]
	\centering{%
\JO[ {%
		\begin{tabular}{rcccccc}
			LZ77 & {\tt 1} & {\tt 2} & {\tt 3} & {\tt 4} & {\tt 5} & {\tt 6} \\
			\toprule
			{\color{LZNode}Factor} &
			{\color{LZNode}\tt a} &
			{\color{LZNode}\tt a} &
			{\color{LZNode}\tt b} &
			{\color{LZNode}\tt aaba} &
			{\color{LZNode}\tt ba} &
			{\color{LZNode}\tt \$}
			\\\midrule
			{\color{takeLeaf}Coding} &
			{\color{takeLeaf}\tt a} &
			{\color{takeLeaf}\tt 1,1} &
			{\color{takeLeaf}\tt b} &
			{\color{takeLeaf}\tt 1,4} &
			{\color{takeLeaf}\tt 3,2} &
			{\color{takeLeaf}\tt \$}
			\\\bottomrule
		\end{tabular}
		\hspace{1em}
		\begin{tabular}{rcccccc}
			LZ78 & {\tt 1} & {\tt 2} & {\tt 3} & {\tt 4} & {\tt 5} & {\tt 6}\\
			\toprule
			{\color{LZNode}Factor} &
						{\color{LZNode}\tt a} &
						{\color{LZNode}\tt ab} &
						{\color{LZNode}\tt aa} &
						{\color{LZNode}\tt b} &
						{\color{LZNode}\tt aba} &
						{\color{LZNode}\tt \$} 
			\\\midrule
			{\color{takeLeaf}Coding} &
						{\color{takeLeaf}\tt a} &
						{\color{takeLeaf}\tt 1,b} &
						{\color{takeLeaf}\tt 1,a} &
						{\color{takeLeaf}\tt b} &
						{\color{takeLeaf}\tt 2,a} &
						{\color{takeLeaf}\tt \$} 
			\\\bottomrule
		\end{tabular}
		\caption{We parse the text \exampleString{} by both factorizations.
		The coding represents a fresh factor by a single character, and a referencing factor by a tuple with two entries.
		For	LZ77, this tuple consists of the referred position and the number of characters to copy.
		For LZ78, it consists of the referred index and a new character.
	}
} ]{%
\begin{description}
	\item[LZ77]
		\begin{tabular}{r*{7}{c}}
			& 1 & 2 & 3 & 4 & 5 & 6 & 7 \\
			\toprule
			{\color{LZNode}Factor} &
			{\color{LZNode}\tt a} &
			{\color{LZNode}\tt aa} &
			{\color{LZNode}\tt b} &
			{\color{LZNode}\tt aba} &
			{\color{LZNode}\tt aaba} &
			{\color{LZNode}\tt aba} &
			{\color{LZNode}\tt \$}
			\\\midrule
			{\color{takeLeaf}Coding} &
			{\color{takeLeaf}\tt a} &
			{\color{takeLeaf}\tt 1,2} &
			{\color{takeLeaf}\tt b} &
			{\color{takeLeaf}\tt 3,3} &
			{\color{takeLeaf}\tt 2,4} &
			{\color{takeLeaf}\tt 3,3} &
			{\color{takeLeaf}\tt \$}
			\\\bottomrule
		\end{tabular}
	\item[classic-LZ77]
		\begin{tabular}{r*{5}{c}}
			& 1 & 2 & 3 & 4 & 5\\
			\toprule
			{\color{LZNode}Factor} &
			{\color{LZNode}\tt a} &
			{\color{LZNode}\tt aab} &
			{\color{LZNode}\tt abaa} &
			{\color{LZNode}\tt abaab} &
			{\color{LZNode}\tt a\$} 
			\\\midrule
			{\color{takeLeaf}Coding} &
			{\color{takeLeaf}\tt a} &
			{\color{takeLeaf}\tt 1,2,b} &
			{\color{takeLeaf}\tt 3,3,a} &
			{\color{takeLeaf}\tt 5,4,b} &
			{\color{takeLeaf}\tt 1,1,\$} 
			\\\bottomrule
		\end{tabular}
\vspace{1em}
				\item[LZ78]
					\begin{tabular}{r*{8}{c}}
						& 1 & 2 & 3 & 4 & 5 & 6 & 7 & 8\\
			\toprule
			{\color{LZNode}Factor} &
						{\color{LZNode}\tt a} &
						{\color{LZNode}\tt aa} &
						{\color{LZNode}\tt b} &
						{\color{LZNode}\tt ab} &
						{\color{LZNode}\tt aaa} &
						{\color{LZNode}\tt ba} &
						{\color{LZNode}\tt aba} &
						{\color{LZNode}\tt \$} 
			\\\midrule
			{\color{takeLeaf}Coding} &
						{\color{takeLeaf}\tt a} &
						{\color{takeLeaf}\tt 1,a} &
						{\color{takeLeaf}\tt b} &
						{\color{takeLeaf}\tt 1,b} &
						{\color{takeLeaf}\tt 2,a} &
						{\color{takeLeaf}\tt 3,a} &
						{\color{takeLeaf}\tt 4,a} &
						{\color{takeLeaf}\tt \$} 
			\\\bottomrule
		\end{tabular}
\end{description}
\caption{We parse the text \exampleString{} by both factorizations.
		The coding represents a fresh factor by a single character, and a referencing factor by a tuple with two entries.
		For	LZ77, this tuple consists of the referred position and the number of characters to copy.
		For classic-LZ77, the tuple additionally contains a new character.
		For LZ78, it consists of the referred index and a new character.
	}
}%
	}%
\label{figParsing}
\end{figure}

\JO{\SubSection{Scaffold of the Algorithms}}
Common to our LZ77- and LZ78-factorization algorithms is the traversal of the compressed suffix tree.
In more detail, they share a common framework, which we describe in the following by introducing some new keywords:

{\bf Witnesses. }
Witnesses are \emph{internal} nodes that act as signposts for finding (LZ77) the referred position or (LZ78) the referred index of a factor.
The \emph{number of witnesses}~\numWitness{} is at most the number of \emph{referencing} factors~\numRefs{}.
We will enumerate the witnesses from $1$ to $\numWitness$ by a bit vector~\bv{W} on the BP of \ST{} with a $\rank[1]$-support.
So each witness has, along with its pre-order number, a so-called \intWort{witness id} (its \bv{W}-rank).

{\bf Passes. }
Like the LZ77 algorithm in~\cite{lzciss}, we divide our algorithms in several passes.
In a pass, we visit the leaves of \ST{} in text position order.
This is done by using $\smallestleaf$ and then calling $\nextleaf$ successively.
The passes differ in how a leaf is processed.
While processing a leaf~$\ell$, we want to access \lab[\ell].
We can track the label of the current leaf with a counter variable, since we start at the leaf with the label $1$.

{\bf Corresponding Leaves. }
We say that a leaf~$\ell$ \intWort{corresponds to} the factor~$f$ if $\lab[\ell]$ is the factor position of~$f$.
During a pass, we keep track of whether a visited leaf corresponds to a factor.
To this end, for each leaf~$\ell$ corresponding to a factor~$f$, we compute the length of~$f$ while processing~$\ell$.
This length tells us the number of leaves after $\ell$ (in \emph{text order}) that do not correspond to a factor.
By noting the next corresponding leaf, 
we know whether the current leaf is corresponding to a factor --- remember that a pass selects leaves successively in \emph{text order}, 
and \smallestleaf{} is always corresponding to the first factor.

{\bf Output Space. }
Given \ST{} and $\psi$, we analyze our algorithms for both factorizations with respect to time and working space.
\JO[We assume that the output can be streamed sequentially in the order of the factor indices. Therefore, we do \emph{not} analyze the output space.]{%
We analyze both factorizations under the assumptions that either the output has to be stored explicitly in RAM, 
or that the algorithm must stream the output sequentially.  }

\JO[\textbf{Loaded Data Structures in RAM\@. }]{\SubSection{Loaded Data Structures in RAM}}
We need the $\psi$-array taking $\Oh{n \lg \sigma}$ bits, and \ST{}'s topology \JO[plus some rank- and select-support data structures]{} consuming $4n + \oh{n}$ bits.
\JO{%
	We spend $\oh{n}$ additional bits to create a $\rank[{\tt ()}]$-support, a $\select[{\tt ()}]$-support, and a $\rank[{\tt (}]$-support on the BP sequence of \ST{} for ranking and selecting leaves and internal nodes separately 
	(nodes and leaves in the BP representation are given by the sequence \bsq{{\tt (}} and \bsq{{\tt ()}}, respectively).
}
Our algorithms do \emph{not} access the text~$T$.

\Section{LZ77}
\JO[%
Given that the above stated data structures are loaded into RAM,
we show that the LZ77 factorization can be computed with $2n + z \lg n + \oh{n}$ bits of working space.
]{%
Given that the above stated data structures are loaded into RAM,
we show that the LZ77 factorization can be computed with $(1+\Entr{n,z})n + z \lg n + \oh{n}$ additional bits of working space when streaming.
It is easy to show that we can store the output in RAM with additionaly $z \lg n$ bits.}

\JO{%
\begin{algorithm}[tb]
\begin{algorithmic}[1]
	\STATE $\ell \gets \smallestleaf$
	\STATE $p \gets 1$ \COMMENT{tracks next leaf corresponding to a factor}
\REPEAT
	\STATE $v \gets \parent[\ell]$
	\WHILE{$v \not= \rootnode$}
		\IF[already visited?]{$\bv{V}[v] = 1$}
		\IF[if the current leaf corresponds to a factor] {$\lab[\ell] = p$} %
				\STATE $\bv{W}[v] \gets 1$ \COMMENT{then this node is a witness}
\STATE $p \gets p +  \strdepth(v)$ \COMMENT{determine next starting factor}
			\ENDIF
			\BREAK \COMMENT{on finding a visited node we stop}
		\ENDIF
		\STATE $\bv{V}[v] \gets 1$ \COMMENT{visit the node}
		\STATE $v \gets \parent[v]$ \COMMENT{move upwards}
	\ENDWHILE	
	\IF[$\ell$ corresponds to a fresh factor]{$v = \rootnode$} \label{alglineFreshFactor}
	\INCR $p$ \COMMENT{factor is a single character}
	\ENDIF
\STATE $\ell \gets \nextleaf[\ell]$
\UNTIL{$\ell = \smallestleaf$}
\end{algorithmic}
\caption{LZ77 \Cref{it77pass1}}
\label{algoLZ77P1}
\end{algorithm}
}%

\JO{%
\begin{algorithm}[t!]
\begin{algorithmic}[1]
	\STATE $\bv{V}.\clear$
	\STATE $\bv{W}.\addRank$
	\STATE $p \gets 1$ \COMMENT{tracks next leaf corresponding to a factor}
	\STATE $\numWitness \gets \bv{W}.\rank[1](n)$
	\STATE $W \gets \text{array of size~} \numWitness \lg n$ \COMMENT{maps witness ids to text positions}
	\STATE $\ell \gets \smallestleaf$
	\REPEAT
	\STATE $v \gets \parent[\ell]$
	\WHILE{$v \not= \rootnode$}
	\IF%
	[{Invariant: $\bv{V}[v] = 1 \wedge p = \lab[\ell] \Rightarrow \bv{W}(v) = 1$}]%
	{$\bv{V}[v] = 1$} 
	
	\IF[$\ell$ corresponds to a factor]{$\lab[\ell] = p$}  
			\STATE output text position $W[\bv{W}.\rank[1](v)]$
			\STATE output factor length $\strdepth[v]$
			\STATE $p \gets p +  \strdepth(v)$ \COMMENT{determine next starting factor}
		\ENDIF
	\BREAK
	\ENDIF
	\IF{$\bv{W}[v] = 1$} 
	\STATE $W[\bv{W}.\rank[1](v)] \gets \lab[\ell]$
	\ENDIF
		\STATE $\bv{V}[v] \gets 1$
	\STATE $v \gets \parent[v]$
  \ENDWHILE

  \IF[We are currently processing a leaf of a fresh factor]{$\lab[\ell] = p$} 
  \STATE output character $\head[\ell]$
	  \STATE output factor length $1$
		\INCR $p$
  \ENDIF
  \STATE $\ell \gets \nextleaf[\ell]$
  \UNTIL{$\ell = \smallestleaf$}
\end{algorithmic}
\caption{LZ77 \Cref{it77pass2}}
\label{algoLZ77P2}
\end{algorithm}
}%

{\bf LZ77 Passes. }
Common to all passes is the following procedure:
For each visited leaf~$\ell$, we perform a leaf-to-root traversal, i.e., we visit every node on the path from~$\ell$ to~$\rootnode$.
But we visit every node at most once, i.e., we stop the leaf-to-root traversal on visiting an already visited node.
Therefore, we create a bit vector~\bv{V} with which we mark a visited node.
This bit vector is cleared before a pass starts.
Since \ST{} contains at most $n-1$ internal nodes, a pass can be conducted in linear time.

We perform two passes:
\begin{passes}
\JO[\crampedItems{}]{}
\item create $\bv{W}$ in order to determine the witnesses\JO{~(see \Cref{algoLZ77P1})}, and \label{it77pass1}
\item stream the output by using an array mapping witness ids to text positions\JO{~(see \Cref{algoLZ77P2})}. \label{it77pass2}
\end{passes}

{\bf \Cref{it77pass1}. }
We follow the approach from~\cite{lzciss}.
Determining the witnesses is done in the following way:
Reaching the root from a leaf corresponding to a factor \JO{(while visiting only non-marked nodes)} means that we found a fresh factor.
Otherwise, assume that we visit an already visited node~$u \not= \rootnode$ from a leaf~$\ell$. 
If $\ell$ corresponds to a factor~$f$, $u$ \emph{witnesses} the referred position of~$f$.
This means that there is a suffix starting before~\lab[\ell] having a prefix 
equal to the string read from the edge labels on the path from the root to $u$.
Moreover, $u$ is the lowest node in the set\JO{~$\menge{ \text{LCA of } \ell \text{~and~} \ell' : \lab[\ell'] < \lab[\ell]}$ } comprising the lowest common ancestors of~$\ell$ with all already visited leaves.
So the factor corresponding to~$\ell$ has to refer to a text position coinciding with the label of a leaf belonging to $u$'s subtree.
In order to find the referred position in the next pass, we mark $u$ in~\bv{W}.
Additionally, we compute the length of~$f$ with $\strdepth[u]$, and note the next factor position.

\begin{figure}[t]
	\JO[ {%
	\centering{%
		\begin{tabular}{*{2}{m{0.5\linewidth}}}
			\includegraphics{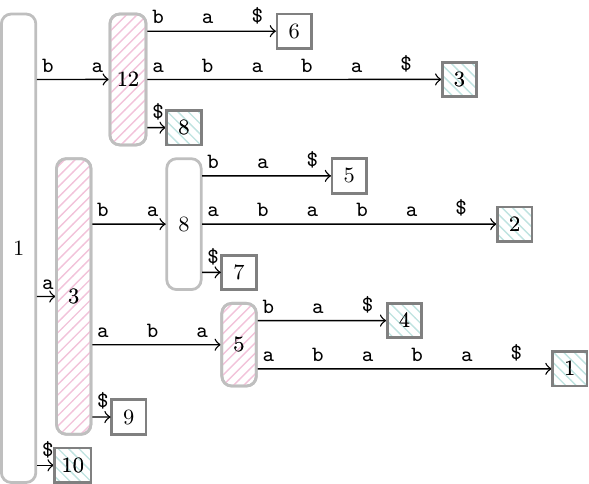}
		&
				 \begin{tabular}{r*{5}{c}}
				 	pre-order number & {\tt 1} & {\tt 3} & {\tt 5} & {\tt 8} & {\tt 12} \\
				 	\toprule
				 	$\bv{W}$         & {\tt 0} & {\tt 1} & {\tt 1} & {\tt 0} & {\tt 1}
				 	\\\bottomrule
				 	\\
				 \end{tabular}
				 \begin{tabular}{rlll}
				 	witness id & {\tt 1} & {\tt 2} & {\tt 3} \\
				 	\toprule
				 	$W$        & {\tt 1} & {\tt 1} & {\tt 3}
				 	\\\bottomrule
				 \end{tabular}
			 \end{tabular}
	}
} ]{%
	\centering{%
		\begin{tabular}{*{2}{m{0.5\linewidth}}}
	\includegraphics{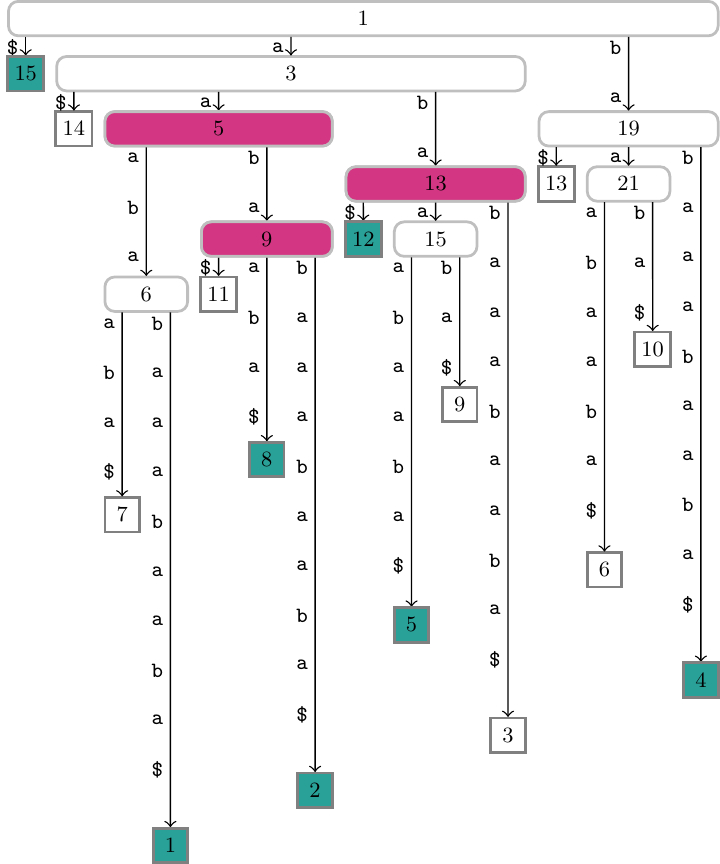}
		&
		\tabcolsep=0.2em
				 \begin{tabular}{r*{9}{c}}
					 pre-order & {\tt 1} & {\tt 3} & {\tt 5} & {\tt 6} &  {\tt 9} & {\tt 13} & {\tt 15} & {\tt 19} & {\tt 21} \\
				 	\toprule
					$\bv{W}$         & {\tt 0} & {\tt 0} & {\tt 1} & {\tt 0} & {\tt 1}   & {\tt 1} & {\tt 0}   & {\tt 0} & {\tt 0} 
				 	\\\bottomrule
				 	\\
				 \end{tabular}
				 \begin{tabular}{rlll}
				 	witness id & {\tt 1} & {\tt 2} & {\tt 3} \\
				 	\toprule
					$W$        & {\tt 1} & {\tt 2} & {\tt 3}
				 	\\\bottomrule
				 \end{tabular}
			 \end{tabular}
	}
}%
	\caption{%
		Our LZ77 algorithm determines the witness nodes and the leaves corresponding to factors in \Cref{it77pass1}.
		Considering our running example $T = \exampleString$, the witness nodes are the nodes with the pre-order numbers 
		\JO[$3,5$, and $12$]{$5, 10$, and $14$},
		and the leaves corresponding to factors have the labels \JO[$1,2,3,4,8,$ and $10$]{$1,2,4,5,8,12,$ and $15$}.
		Each witness~$w$ is the lowest ancestor of a leaf corresponding to a factor~$f$ with the property that the referred position of~$f$
		is the label of a leaf contained in $w$'s subtree.
		For instance, the leaf corresponding to the \JO[$5$]{$4$}-th factor has the label~\JO[$8$]{$5$}. Its witness has pre-order number \JO[$12$]{$13$}, leading to 
		the leaf with the label~\JO[$3$]{$3$}. So the referred position of the \JO[$5$]{$4$}-th factor is~$3$.
		The length of the \JO[$5$]{$4$}-th factor is the string depth of its witness. 
		We show $W$ and $\bv{W}$ after \Cref{it77pass2}. In this example, \JO[$\numWitness = \numRefs = 3$]{$\numWitness = 3$ and $\numRefs = 4$}.
	}
\label{figWitness}
\end{figure}

After this pass, we have determined the \numWitness{} witnesses by the \bsq{1}s stored in $\bv{W}$.
We use the witnesses in the next pass to compute the referred positions~(see \Cref{figWitness}).

{\bf \Cref{it77pass2}. }
We clear $\bv{V}$, create a rank-support on $\bv{W}$ 
and allocate an array $W$ consuming $\numWitness \lg n$ bits.
We use $W$ to map a witness id to a text position\JO{~(referred positions)}.
Having this array as a working space, $W[w]$ becomes the label of the leaf
from which we visited the witness~$w$ in the first place.
So we find the referred position of a referencing factor~$f$ in $W[w]$ when visiting~$w$ again from a different leaf corresponding to $f$.
The length of~$f$ is the string depth of $w$. 
Since fresh factors consist of single characters, we can output a fresh factor by applying the $\head$-function to its corresponding leaf.

\JO{%
{\bf Compressing \bv{W}. }
Instead of directly marking nodes in \bv{W}, we can allocate $z \lg n$ bits (we can use the space later for the array~$W$) 
storing the pre-order numbers of all witnesses
during \Cref{it77pass1}.
Afterwards, we can create a compressed bit vector with constant time rank-support, representing \bv{W}.
}%

\JO{%
\SubSection{Trade-Off Variant}
We can reduce $W$ to $\varepsilon z \lg z$ by performing both passes $1/\varepsilon$ times.
Therefore, we prematurely stop \Cref{it77pass1} after counting $\varepsilon$~many witnesses.
We store the label~$j$ of the last visited leaf in order to resume \Cref{it77pass1} after outputting the found factors.
Since \numWitness{} is now~$\varepsilon$, we need not modify \Cref{it77pass2}.
When running \Cref{it77pass1} again to capture the next $\varepsilon$ witnesses (some may be the same), 
we suppress the marking in $\bv{W}$ when visiting leaves 
corresponding to referencing factors whose factor positions are at most $j$.

Since we run both passes $1/\varepsilon$ times we get $\Oh{n/\varepsilon}$ time overall.
}%

\JO{%

\SubSection{Classic LZ77 factorization}
The LZ77 and the classic-LZ77 factorization differ in the fact that a factor introduces always a new character at its end in the classic flavor.
We can easily adopt our LZ77 factorization algorithm to the classic flavor.
To this end, when processing a corresponding leaf, we skip additionally to the string depth of its witness one character, 
making the currently processed factor longer by one character.
Like in the LZ78 part, we can get the new character of a referencing factor with factor index~$x$ by accessing the leaf~$\ell'$ whose label
is one text position smaller than the label of the leaf corresponding to the ($x+1$)-th factor.
We conducted the classic LZ77 factorization on our running example in \Cref{figClassic}.

We additionally have to check in \Cref{algoLZ77P1} line~\ref{alglineFreshFactor} that we reach the root from a 
corresponding leaf. This is no longer an invariant: If this condition does not hold, then the previous factor consumed 
the new character that would be used as a fresh factor in our standard variant.

\begin{figure}[t]
	\centering{%
		\begin{tabular}{*{2}{m{0.5\linewidth}}}
			\includegraphics{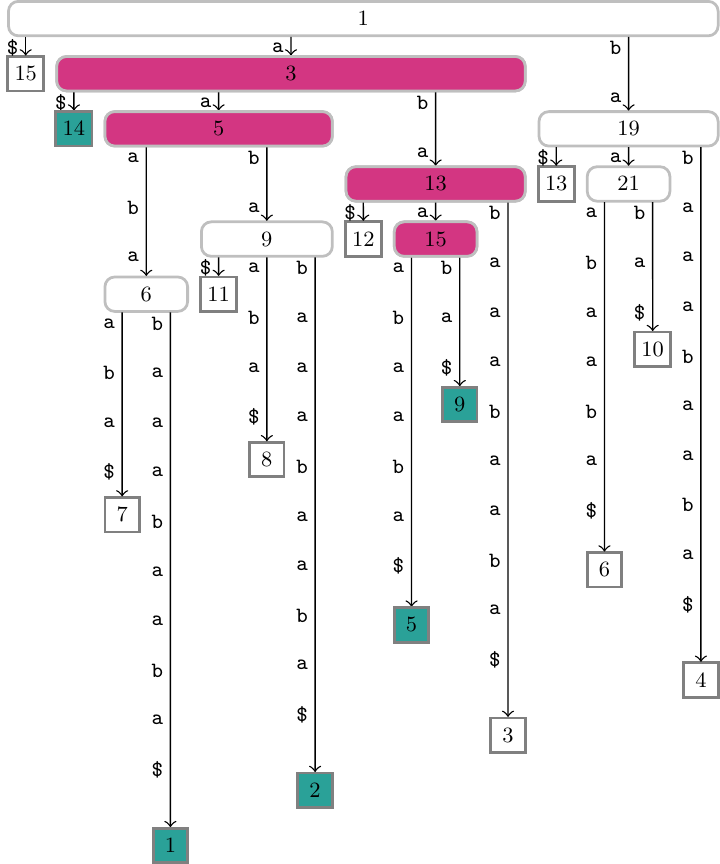}
		&
		\tabcolsep=0.2em
				 \begin{tabular}{r*{9}{c}}
					 pre-order       & {\tt 1} & {\tt 3} & {\tt 5} & {\tt 6} &  {\tt 9} & {\tt 13} & {\tt 15} & {\tt 19} & {\tt 21} \\
				 	\toprule
					$\bv{W}$         & {\tt 0} & {\tt 1} & {\tt 1} & {\tt 0} & {\tt 0}   & {\tt 1} & {\tt 1}   & {\tt 0} & {\tt 0} 
				 	\\\bottomrule
				 	\\
				 \end{tabular}
				 \begin{tabular}{r*{4}{c}}
					 witness id & {\tt 1} & {\tt 2} & {\tt 3} & {\tt 4} \\
				 	\toprule
					$W$        & {\tt 1} & {\tt 1} & {\tt 3} & {\tt 5}
				 	\\\bottomrule
				 \end{tabular}
			 \end{tabular}
	}
	\caption{%
		Applying the LZ77-classic algorithm to our running example $T = \exampleString$, the witness nodes have the pre-order numbers $3,5,13,$ and $15$,
		and the leaves corresponding to factors have the labels $1,2,5,9,$ and $14$.
		In this example, we have $\numWitness = \numRefs = 4$.
	}
\label{figClassic}
\end{figure}

}%

\Section{LZ78}
A natural representation of the LZ78 factors is a trie, the so-called \intWort{LZ trie}.
Each node in the trie represents a factor and is labeled by its index.
If the $x$-th factor refers to the $y$-th factor, 
then there is a node~$u$ having a child~$v$ such that
$u$ and $v$ have the unique labels $y$ and $x$, respectively. 
The edge~$(u,v)$ is labeled by the last character of the $x$-th factor
(the newly introduced character).
A node with the label~$x$ is the child of the root iff the $x$-th factor is a fresh factor.

The LZ trie representation is used in the \JO[algorithm]{algorithms} presented below. 
\JO[%
By building the LZ trie topology on \ST{}, 
our streaming algorithm computes the factorization with $5n + z \lg z + \oh{n}$ additional bits of working space.
]{%
While the streaming algorithm computes the topology of the LZ trie,
the storing variant explicitly creates the LZ trie for querying.
We can compute the factorization with $5n + z \lg z + \oh{n}$ additional bits of working space when streaming the output, 
or with $6n + z (\lg \sigma + \lg z + 3) + \oh{n}$ additional bits of working space when storing the LZ trie explicitly, 
labeling each node of the trie by a factor index.
}%

\begin{figure}[t]
\JO[ {%
	\begin{tabular}{@{}*{2}{m{0.45\textwidth}}}
	\centering{%
		\includegraphics[scale=1.0]{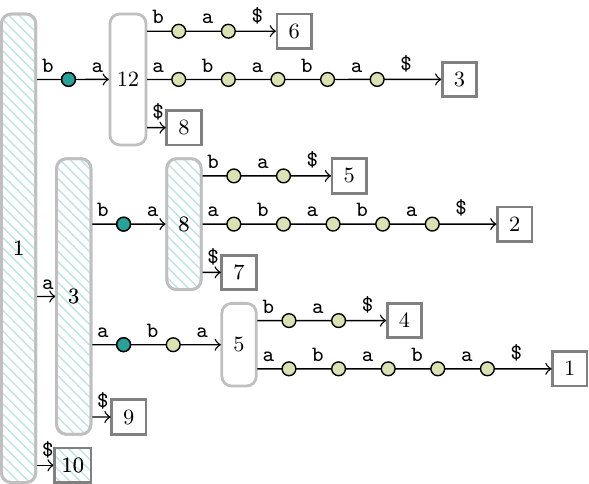}
	}
	&
	\begin{tabular}{r*{10}{c}}
		 $\SA$ & {\tt 10} & {\tt 9} & {\tt 1} & {\tt 4} & {\tt 7} & {\tt 2} & {\tt 5} & {\tt 8} & {\tt 3} & {\tt 6}
		 \\\toprule
		 $\bv{C}$ & {\tt 0} & {\tt 0} & {\tt 0} & {\tt 1} & {\tt 0} & {\tt 0} & {\tt 0} & {\tt 0} & {\tt 0} & {\tt 1}
	 	\\\bottomrule
	 		\end{tabular}

				\begin{tabular}{r*{5}{c}}
					\\
					pre-order number & {\tt 1} & {\tt 3} & {\tt 5} & {\tt 8} & {\tt 12} \\
					\toprule
					$\bv{V}$         & {\tt 0} & {\tt 1} & {\tt 0} & {\tt 1} & {\tt 0} \\
					\midrule
					$\bv{W}$         & {\tt 0} & {\tt 1} & {\tt 1} & {\tt 1} & {\tt 1}
					\\\bottomrule
				\end{tabular}
			\end{tabular}
} ]{%
	\begin{tabular}{@{}*{2}{m{0.5\textwidth}}}
	\hspace{-2em}
	\centering{%
		\includegraphics[scale=1.0]{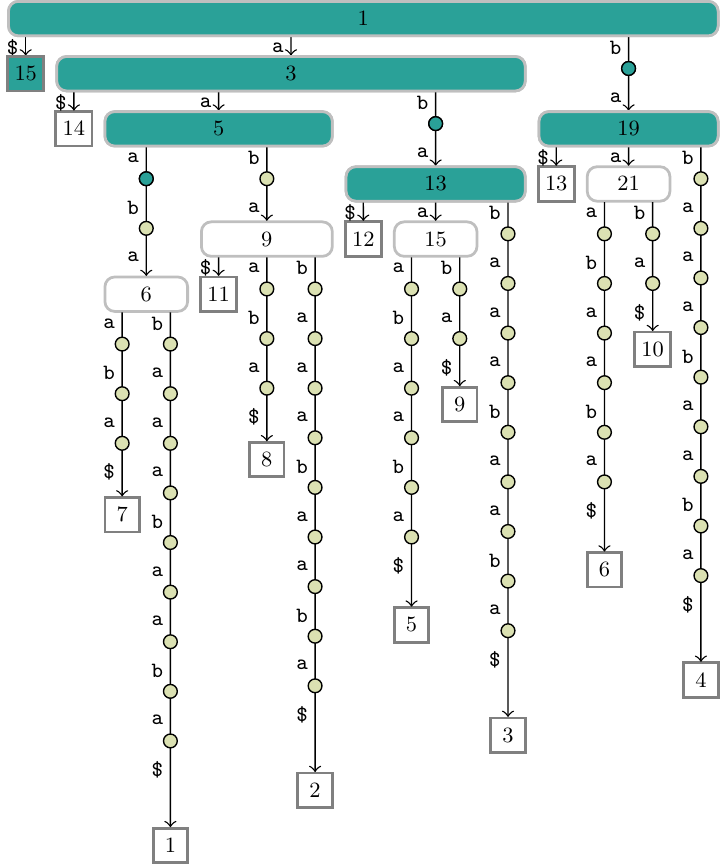}
	}
	&
	\tabcolsep=0.2em
	{\fontsize{10}{10}\selectfont{}
	\begin{tabular}{r*{15}{c}}
		$\SA$       & {\tt 15} & {\tt 14} & {\tt 7} & {\tt 1} & {\tt 11} & {\tt 8} & {\tt 2} & {\tt 12} & {\tt 5} & {\tt 9} & {\tt 3} & {\tt 13} & {\tt 6} & {\tt 10} & {\tt 4}
		 \\\toprule
		 $\bv{C}$   & {\tt 0}  & {\tt 0}  & {\tt 1} & {\tt 0} & {\tt 0}  & {\tt 0} & {\tt 0} & {\tt 0}  & {\tt 0} & {\tt 0} & {\tt 0} & {\tt 0}  & {\tt 0} & {\tt 0}  & {\tt 0} 
	 	\\\bottomrule
	 		\end{tabular}
		}

	\begin{tabular}{r*{9}{c}}
					\\
					pre-order & {\tt 1} & {\tt 3} & {\tt 5} & {\tt 6} &  {\tt 9} & {\tt 13} & {\tt 15} & {\tt 19} & {\tt 21}
					\\	\toprule
					$\bv{V}$         & {\tt 0} & {\tt 1} & {\tt 1} & {\tt 0} & {\tt 0} &  {\tt 1} & {\tt 0} & {\tt 1} & {\tt 0}
\\ \midrule
					$\bv{W}$         & {\tt 0} & {\tt 1} & {\tt 1} & {\tt 1} & {\tt 0} &  {\tt 1} & {\tt 0} & {\tt 1} & {\tt 0}
					\\\bottomrule
				\end{tabular}
			\end{tabular}
}%
	\caption{We can get the suffix trie (conceptually) by exchanging every ST edge~$e$ with $c(e)-1$ new suffix trie nodes superimposing~$e$.
	These new suffix trie nodes are the small rounded nodes in the depicted tree.
	They represent the implicit suffix trie nodes, while the remaining ST nodes represent the explicit suffix trie nodes.
	Dark colored and hatched nodes represent the nodes of the LZ78 trie.
	We show $\bv{C}$, $\bv{V}$, and $\bv{W}$ after \Cref{it78pass1}. %
	}
\label{figLZTrie}
\end{figure}

\JO[\textbf{Superimposition. }]{\SubSection{Superimposition}}
The main idea is the superimposition of the suffix trie on the suffix tree, borrowed from \citet{Nakashima2015CLT}:
The LZ trie is a connected subgraph of the suffix trie containing its root~(see \Cref{figLZTrie}).
Regarding the suffix tree, the LZ nodes are either already represented by an \ST{} node (explicit), or lie on an \ST{} edge (implicit).
To ease explanation, we identify each edge~$e=(u,v)$ of \ST{} uniquely with its ending node~$v$, i.e.,
we implicitly convert between the edge~$e$ and its in-going node~$v$ (each node except \rootnode{} is associated with an edge).
In order to address all LZ nodes, we keep track of how far an edge on the suffix tree got explored during the parsing.
To this end, for an edge~$e=(u,v)$, we define the \intWort{exploration counter} $0 \le n_v \le c(e)$ storing how far $e$ is explored.
If $n_v = 0$, then the factorization has not (yet) explored~$e$,
whereas $n_v = c(e)$ tells us that we have already reached~$v$.
Unfortunately, storing $n_v$ in an integer array for all edges costs us $2n \lg n$ bits.

Our idea is to choose different representations of the exploration counters dependent on the state (not, partially or fully explored) and 
the number of descendants of a node. 
First, we mark the fully explored edges in a bit vector~\bv{V} (dynamically) such that we do not need to store their exploration counters.
Further, we do not represent $n_v$ for a node~$v$ with parent~$u$ until $n_u$ got fully explored.
Now let us focus on the rest of the nodes.
We classify each node~$v$ based on the number of descendants of~$v$, 
and select an explicit representation if this number is large, otherwise we maintain~$n_v$ implicitly.
The classification of~$v$ is based on the following definitions borrow from~\cite{sadakane06squeezing}:
If $v$'s subtree has at most $\lg n$ nodes, we call $v$ \intWort{micro}.
If $v$ is not a micro node, but all its children are micro, then we call~$v$ a \intWort{jump node}.
So a subtree rooted at a jump node contains at least $\lg n$ nodes, and
the subtrees rooted at different jump nodes are pairwise disjoint (they do not share a node).
This means that there are at most $n / \lg n$ jump nodes.
We mark each jump node~$v$ in a bit vector~\bv{J}, and store $n_v$ in an integer array~$J$.
The array~$J$ has at most $n / \lg n$ entries and therefore consumes at most $(n / \lg n) \lg n = n$ bits.
Let us consider a node~$v$ that is not micro and whose in-going edge did get partially explored.
If~$v$ is a jump node, then we look-up~$n_v$ in~$J$.
Otherwise, $v$ has at least one descendant that is a jump node.
Since $v$'s in-going edge is not fully explored, the exploration counters of all edges in the subtree rooted at~$v$ are zero.
So we can abuse an exploration counter of a jump node belonging to this subtree to represent~$n_v$ until $n_v$ gets full.
For instance, we can always use the leftmost jump node of~$v$ that can be accessed by 
$\bv{J}.\select[1](\bv{J}.\rank[1](v)+1)$.

The exploration counters of the micro nodes are maintained implicitly by a bit vector marking visited corresponding leaves:
During a pass, when exploring a new factor on the in-going edge of a micro node~$v$, we mark the currently accessed leaf (which will always be a leaf in the subtree rooted at $v$) in a bit vector~\bv{C}.
By applying \popcount{} to \bv{C}, we can count how many leaves had been accessed belonging to the subtree rooted at~$v$.
This number is exactly~$n_v$.
We can compute $n_v$ in constant time, since there are at most $\lg n$ leaves in the subtree rooted at~$v$.
After fully exploring the edge of~$v$, we clear the area in \bv{C} belonging to the leaves contained in $v$'s subtree.
By doing so, the counter~$n_u$ of every micro child~$u$ of~$v$ is reset.

\JO{Dividing suffix tree nodes in explored/unexplored and not-micro/micro creates two boundaries, as illustrated in \Cref{figBoundaries}.}

Applying this procedure during a pass, we can determine the fully explored edges and 
collect $n_v$~of each node~$v$ whose in-going edge got partially explored 
(by the definition of the jump nodes, and since we clear parts of $\bv{C}$ after full exploration).

\JO{\SubSection{Streaming Variant}}
We do two passes:
\begin{passes}
\JO[\crampedItems{}]{}
\item create \bv{W} so we can address the witnesses\JO{~(see \Cref{algoLZ78P1})}, and \label{it78pass1}
\item stream the output by using a helper array mapping witness ids to factor indices\JO{~(see \Cref{algoLZ78P2})}. \label{it78pass2}
\end{passes}

\newcommand{\findedge}   [1][]{\UnaryOperator[#1]{\textsl{find\_edge}}}
\newcommand{\explore}   [1][]{\UnaryOperator[#1]{\textsl{explore}}}

\JO{%
\begin{algorithm}[t]
\begin{algorithmic}[1]
	\REQUIRE leaf $\ell$
\STATE {$v \gets \rootnode$}
  \STATE {$d \gets 0$} \COMMENT{node depth}
  \REPEAT [find first edge on path from root to $\ell$ that is not fully explored]
  	\INCR $d$
	\STATE {$v \gets \levelanc[\ell, d]$}
	\UNTIL {$v = \ell$ or $\bv{V}[v] = 0$}
	\STATE {$u \gets \parent[v]$} \COMMENT{new factor is on the edge $(u, v)$}
	\RETURN $(u,v)$
\end{algorithmic}
\caption{Function \findedge{} finds the first edge $(u,v)$ on the path from the root to $\ell$ that is not yet fully explored.}
\label{algoLZ78findedge}
\end{algorithm}

\begin{algorithm}[t!]
\begin{algorithmic}[1]
	\ENSURE $\bv{J}[v] \gets 1$ for every jump node~$v$
	\ENSURE $J \gets$ integer array with $n/\lg n$ entries, i.e., $n$ bits.
	\REQUIRE node $v$, $s = \strdepth[v]$
	\IF {$v$ is a jump node}
\STATE {$m \gets J[\bv{J}.\rank[1](v)]$}
\ELSIF {$v$ is not micro}
\STATE {$m \gets J[\bv{J}.\select[1](\bv{J}.\rank[1](v)+ 1)]$}
\ELSE
\STATE $m \gets \popcount(\bv{C}[\lmostleaf[v], \rmostleaf[v]])$ 
  \ENDIF
  \STATE {Let $\ell,\ell'$ be two leaf whose LCA is $v$}
\STATE {$\ell = {(\nextleaf)}^{m+s+1}[\ell]$}
\STATE {$\ell' = {(\nextleaf)}^{m+s+1}[\ell']$}
\IF[edge $(u,v)$ now fully explored] {$\head[\ell] \neq \head[\ell']$}
\STATE {$\bv{V}[v] \gets 1$} \COMMENT{set $v$ as fully explored}
\STATE {\bv{C}.\clear[\lmostleaf[v], \rmostleaf[v]]} \COMMENT{reset the counting so that we can work with the children of $v$}
\ELSIF[edge $(u,v)$ has at least one character unexplored] {$v$ is micro}
\STATE $\bv{C}[\ell] \gets 1$ \COMMENT{increment $n_e$ for the micro node~$v$}
  \ENDIF
  \IF[increment $n_e$ for the jump node~$v$]{$\bv{J}[v] = 1$}
  \INCR{$J[\bv{J}.\rank[1](v)]$}  
  \ELSIF{$v$ is not micro}
  \INCR{$J[\bv{J}.\select[1](\bv{J}.\rank[1](v)+ 1)]$}
  \ENDIF
  \RETURN{$m$}
\end{algorithmic}
\caption{Function \explore{} post-increments $n_v$, returning its previous value.}
\label{algoLZ78explore}
\end{algorithm}

\begin{algorithm}[t]
\begin{algorithmic}[1]
\STATE {$\ell \gets \smallestleaf$} 
\REPEAT
\STATE{$(u,v) \gets \findedge[\ell]$}
\STATE {$s \gets \strdepth[u]$}
\IF {$v = \ell$} 
\STATE {$\ell = {(\nextleaf)}^{s+1}[\ell]$}
	\CONTINUE
\ENDIF
	\STATE {$\bv{W}[v] \gets 1$} \COMMENT{$v$ is a witness}
	\STATE{$m \gets \explore[v,\ell] $} \COMMENT{$m \gets n_v; n_v \gets n_v + 1$}
	\STATE {$\ell = {(\nextleaf)}^{m+s+1}(\ell)$}
\UNTIL{$\ell = \smallestleaf$}
\end{algorithmic}
\caption{LZ78 \Cref{it78pass1}}
\label{algoLZ78P1}
\end{algorithm}

\begin{algorithm}[t]
\begin{algorithmic}[1]
	\STATE $\bv{W}.\addRank$
	\STATE $\numWitness \gets \bv{W}.\rank[1](n)$
	\STATE{$W \gets \text{array of size~} \numWitness \lg z \text{~bits}$}
	\STATE{$x \gets 0$}
	\STATE {$\ell \gets \smallestleaf$} 
\REPEAT
	\STATE{$(u,v) \gets \findedge[\ell]$}
	\STATE {$s \gets \strdepth[u]$}

	\IF{\bv{W}[v] = 0}
		\IF{$v$ is a child of \rootnode{}}
		\STATE{the $x$-th factor is a fresh factor}
		\ELSE
		\STATE{the $x$-th factor refers to $W[\bv{W}.\rank[1](\parent[\ell])]$ }
		\ENDIF
	\ELSE
		\STATE{the $x$-th factor refers to $W[\bv{W}.\rank[1](v)]$ }
	\ENDIF

	\INCR{x}
	\IF {$v = \ell$} 
		\STATE {$\ell = {(\nextleaf)}^{s+1}[\ell]$}
		\CONTINUE
	\ENDIF
	\STATE{$m \gets \explore[v,\ell] $} \COMMENT{$m \gets n_v; n_v \gets n_v + 1$}
	\STATE {$\ell = {(\nextleaf)}^{m+s}(\ell)$}
	\STATE{output character $\head[\ell]$ belonging to the ($x-1$)-th factor}
	\STATE {$\ell = \nextleaf(\ell)$}
\UNTIL{$\ell = \smallestleaf$}
\end{algorithmic}
\caption{LZ78 \Cref{it78pass2}}
\label{algoLZ78P2}
\end{algorithm}

}%

\JO{%
\begin{figure}[t]
	\centering{%
		\includegraphics[scale=1.0]{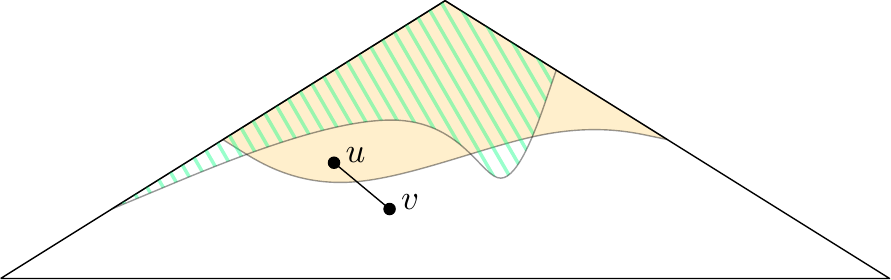}
	}
	\caption{Our LZ78 algorithm divides \ST{} by two boundaries:
		Nodes having at least $\lg n$ nodes in their subtree (hatched upper cone), and
		(partially) explored edges belonging to the LZ trie (colored upper cone).
		The nodes whose exploration counters are stored implicitly are directly below the fringe of the LZ trie.
		During the passes, we always search for an \ST{} edge~$(u,v)$ crossing the boundary of the already discovered part.
	}
\label{figBoundaries}
\end{figure}
}%
We explain the passes in detail, after introducing their commonality and a helpful lemma:

{\bf LZ78 Passes. }
Since referencing factors address factor indices ($z$ options) instead of text positions ($n$ options),
we are only interested in the leaves 
corresponding to a factor. 
Starting with \smallestleaf{}, which corresponds to the first factor,
we can compute the length of the factor corresponding to the currently accessed leaf so that 
we know the distance (in text positions) to the next corresponding leaf.

\begin{lemma}[{\cite[Lemma 4]{lzciss}}]\label{lemma78EdgeCount}
	Let $e=(u,v)$ be an \ST{} edge, and $u$ the parent of the node~$v$.
	Then $n_v \le \min\tuple{c(e),s}$, where $s$ is the number of leaves of the subtree rooted at~$v$.
\end{lemma}

{\bf \Cref{it78pass1}. }
The main goal of this pass is to determine the topology of the LZ trie with respect to the superimposition.
Starting with an LZ trie consisting only of the root, we build the LZ trie successively by filling up the exploration counters.
If the exploration counter of an edge is filled up, we mark its in-going node in the bit vector \bv{V}.

Assume that we visit a leaf~$\ell$.
We want to find the first edge on the path from \rootnode{} to $\ell$ that is either unexplored or partially explored.
By invoking level ancestor queries, we traverse from the root to an edge~$e=(u,v)$, 
where $n_v < c(e)$ and $u$ is (already) represented as a node in the LZ trie.
If $v$ is an internal node with $n_v = 0$, we make $v$ a witness by marking $v$ in~$\bv{W}$ (the idea is that the edge~$e$ is \emph{superimposed} by some LZ nodes).

Regardless that, we add a new factor by incrementing $n_v$.
If the edge~$e$ now got fully explored, we additionally mark~$v$ in $\bv{V}$.
Whether the edge~$e$ got fully explored, can be determined with the $\nextleaf$ function:
First, if $v$ is a leaf, the edge~$(u,v)$ can be explored at most once (by \Cref{lemma78EdgeCount}).
Otherwise, we choose a leaf~$\ell'$ such that the LCA of~$\ell$ and~$\ell'$ is~$v$.
The idea is that \strdepth[v] is the length of the longest common prefix of two suffixes corresponding to two leaves (e.g., $\ell$ and $\ell'$) having $v$ as their LCA\@.
So we can compare the $m$-th character of both respective suffixes 
by applying $\nextleaf$ $m$-times on both leaves before using the \head{}-function.
With $m := \strdepth[u]+n_v+1$ we can check whether the edge~$(u,v)$ got fully explored.
Additionally, we can determine the label of the next corresponding factor.
Although we apply \nextleaf{} as many times as the factor length, we still get linear time overall, because concatenating all factors yields the text~$T$.

{\bf \Cref{it78pass2}. }
This pass is nearly identical to \Cref{it78pass1}.
	We explore the LZ trie nodes again, 
	but this time we already have the witnesses.
	So we keep \bv{W}, but reset the exploration counters and \bv{V}. 

	For finding the referred indices, we create an array $W$ with $\numWitness \lg z$ bits to store a factor index for each witness id.
	The witness ids are determined by \bv{W}. The factor indices are given by a counter variable 
	tracking the number of visited corresponding leaves, i.e., the number of processed factors.

	Assume that we visit the leaf~$\ell$ corresponding to the $x$-th factor, i.e., $\ell$ is the $x$-th visited corresponding leaf.
	Again by level ancestor queries, we determine the edge~$e=(u,v)$ on the path from the root to $\ell$, where $u$ is in \bv{V} and $v$ not.

	If $v$ is an internal node, then $v$ is a witness.
	In this case, we retrieve $y := W[\bv{W}.\rank[1](v)]$.
	If $y$ is defined, then the $x$-th factor refers to the $y$-th factor.

	If $y$ is undefined (i.e., its value has not yet been initialized), or if $v$ is a leaf (i.e., $v = \ell$), 
	then the $x$-th factor is either a fresh factor if $v$ is a child of~\rootnode,
	or the $x$-th factor refers to $W[\bv{W}.\rank[1](\parent[v])]$.

	If $v$ is an internal node, we set $W[\bv{w}.\rank[1](v)] \gets x$, and
	increment $n_v$ (thus exploring the LZ trie like before).
	Like before, when $v$'s in-going edge gets fully explored, we mark $v$ in \bv{V}.

	So far, we can output the referred index of the $x$-th factor, if it exists.
	We get the new character of the $x$-th factor (i.e., the last letter of the factor) by accessing
	the leaf~$\ell'$ that is (with respect to text order) \emph{before} the leaf corresponding to the $(x+1)$-th factor;
	then we can output the new character by $\head[\ell']$.
	
\JO{%
\SubSection{Explicitly storing the LZ trie}
Our goal is to represent the LZ78 factorization by three data structures:
The first one is a balanced parenthesis sequence storing the LZ trie topology.
It is accompanied by an array storing the factor index and an array storing the ending character, for each LZ node.
We can build the balanced parenthesis sequence directly after \Cref{it78pass1}:
To this end, we traverse the suffix tree with a depth first search starting at the root.
While traversing \ST{} we create $c(e)$ LZ nodes for each edge $e=(u,v)$
if $v$ belongs to \bv{V}, or create $n_v$~LZ~nodes otherwise.
We can retrieve the value of $n_v$ or $c(e)$ because of the following:
For the former, we know the exploration counters of all \emph{partial} discovered edges during \Cref{it78pass1}.
For the latter, we can recompute the $c(\cdot)$ values of the \emph{fully} discovered edges with \strdepth{}
without worsening the linear time bound.

During the traversal of \ST{} we can build the $z \lg \sigma$-bits array storing the ending characters:
Assume that we create the LZ nodes $u'$ and $v'$ on the \ST{} edge~$(u,v)$, where $u'$ is the parent of $v'$.
We can obtain the label of the edge $(u',v')$ consisting of a character by invoking \nextleaf{} and \head{}:
Given that $s$ is the string depth of $v'$ in the LZ trie,
applying $\nextleaf$ $s$ times to a leaf~$\ell$ in the subtree rooted at~$v$ returns a leaf~$\ell'$ whose \head[\ell']-value 
is the label in question.

Overall, we can compute the LZ trie in \Oh{n} time, storing its BP representation 
along with the edge labels in $2z + z \lg \sigma + \oh{z}$ bits.

Annotating the LZ nodes with the factor indices cannot be done in linear time without a pre-computation step, which
is topic of this paragraph.
Here, we want to compute a mapping between suffix tree nodes and LZ nodes such that we can access
the LZ node corresponding to the currently explored factor during a pass in constant time.
To this end we create a bit vector~\bv{U} marking suffix tree nodes and a bit vector~\bv{E} marking LZ nodes.
The former marks all suffix tree nodes~$v$ with $n_v > 0$,
the latter marks all LZ trie children of every explicit LZ node.
The last thing we do is adding a $\rank{}$-support to \bv{U}, and a \select{}-support to \bv{E}.
Now assume that there is a suffix tree node~$v$ marked in \bv{U}.
The LZ trie contains a node~$v'$ marked with \bv{E} such that $v' = \bv{E}.\select[1](\bv{U}.\rank[1](v))$ (conceptually $\bv{E}.\rank[1](v') = \bv{U}.\rank[1](v)$).
The node~$v'$ has the following properties:
First, the LZ trie parent of $v'$ is an explicit LZ node that is represented by the suffix tree parent of $v$ ($v$ has a parent since the root is not marked in \bv{U}).
Second, there are $n_v-1$ nodes below $v'$ forming a unary tree\JO{~(also called linear graph or path graph)}.
Due to the balanced parenthesis representation of the LZ trie, the query $v'.\descendant[j]$ 
returning the descendant of $v'$ at depth $\depth[v']+j$ 
for $0 \le j \le n_v-1$
can be answered in constant time.

Now we explain the final pass.
We exchange \Cref{it78pass2} with 
\begin{passes}[resume]
\item store the factor index and the ending character of each LZ node in an array~$W$.\label{it78pass3}
\end{passes}

{\bf \Cref{it78pass3}. }
We create an array $W$ with $z \lg z$ bits to store a factor index.
We recompute \bv{V} and the exploration counters.
We count the current factor index with a variable~$x$.
Assume that we visit the leaf~$\ell$ corresponding to the $x$-th factor.
By level ancestor queries, we find the edge~$e=(u,v)$ in \ST{} on the path from \rootnode{} to $\ell$, where $u$ belongs to $\bv{V}$, but $v$ not.
We set 
\(
	W[\bv{E}.\select[1](\bv{U}.\rank[1](u)).\descendant[n_v]] \gets x.
\)

}%

\subsection*{Acknowledgements}
We thank Veli Mäkinen for outlining us the differences between~\cite{djamalCSTOld} and~\cite{djamalCST}, 
and Johannes Fischer for some helpful comments.
Most parts of this work was done during a visit at the graduate school of information science and technology of the university of Tokyo,
supported by the \emph{Studienwerk für Deutsch-Japanischen Kulturaustausch in NRW e.V}.

\bibliographystyle{IEEEtranN}
\bibliography{ref}

\begin{thebibliography}{20}
\providecommand{\natexlab}[1]{#1}
\providecommand{\url}[1]{#1}
\csname url@samestyle\endcsname
\providecommand{\newblock}{\relax}
\providecommand{\bibinfo}[2]{#2}
\providecommand{\BIBentrySTDinterwordspacing}{\spaceskip=0pt\relax}
\providecommand{\BIBentryALTinterwordstretchfactor}{4}
\providecommand{\BIBentryALTinterwordspacing}{\spaceskip=\fontdimen2\font plus
\BIBentryALTinterwordstretchfactor\fontdimen3\font minus
  \fontdimen4\font\relax}
\providecommand{\BIBforeignlanguage}[2]{{%
\expandafter\ifx\csname l@#1\endcsname\relax
\typeout{** WARNING: IEEEtranN.bst: No hyphenation pattern has been}%
\typeout{** loaded for the language `#1'. Using the pattern for}%
\typeout{** the default language instead.}%
\else
\language=\csname l@#1\endcsname
\fi
#2}}
\providecommand{\BIBdecl}{\relax}
\BIBdecl

\bibitem[Ziv and Lempel(1977)]{Ziv1977uas}
J.~Ziv and A.~Lempel, ``{A Universal Algorithm for Sequential Data
  Compression},'' \emph{{IEEE} Transactions on Information Theory}, vol.~23,
  no.~3, pp. 337--343, 1977.

\bibitem[Ziv and Lempel(1978)]{Ziv1978Coi}
------, ``{Compression of Individual Sequences via Variable-Rate Coding},''
  \emph{{IEEE} Transactions on Information Theory}, vol.~24, no.~5, pp.
  530--536, 1978.

\bibitem[Kosolobov(2015)]{kosoblovLZ}
D.~Kosolobov, ``{Faster Lightweight Lempel-Ziv Parsing},'' in
  \emph{{MFCS}}.\hskip 1em plus 0.5em minus 0.4em\relax Springer Berlin
  Heidelberg, 2015, vol. 9235, pp. 432--444.

\bibitem[Kärkkäinen et~al.(2013)Kärkkäinen, Kempa, and
  Puglisi]{kempaLightLZ}
J.~Kärkkäinen, D.~Kempa, and S.~J. Puglisi, ``{Lightweight Lempel-Ziv
  Parsing},'' in \emph{{Experimental Algorithms}}, 2013, pp. 139--150.

\bibitem[Belazzougui and Puglisi(2016)]{djamalLZ}
D.~Belazzougui and S.~J. Puglisi, ``{Range Predecessor and Lempel-Ziv
  Parsing},'' \emph{SODA}, pp. 2053--2071, 2016.

\bibitem[Belazzougui(2014)]{djamalCSTOld}
D.~Belazzougui, ``{Linear Time Construction of Compressed Text Indices in
  Compact Space},'' in \emph{STOC}.\hskip 1em plus 0.5em minus 0.4em\relax ACM,
  2014, pp. 148--193.

\bibitem[Burrows and Wheeler(1994)]{bwt}
M.~Burrows and D.~J. Wheeler, ``{A Block-sorting Lossless Data Compression
  Algorithm},'' Digital Equipment Corporation, Tech. Rep., 1994.

\bibitem[Belazzougui(2015)]{djamalCST}
D.~Belazzougui, ``{Linear time construction of compressed text indices in
  compact space},'' \emph{ArXiv CoRR}, vol. abs/1401.0936, 2015.

\bibitem[Fischer and Gawrychowski(2015)]{fischer15alphabet}
J.~Fischer and P.~Gawrychowski, ``{Alphabet-Dependent String Searching with
  Wexponential Search Trees},'' in \emph{{CPM}}, 2015, pp. 160--171.

\bibitem[Jansson et~al.(2015)Jansson, Sadakane, and Sung]{Jansson2015LDT}
J.~Jansson, K.~Sadakane, and W.~Sung, ``{Linked Dynamic Tries with Applications
  to {LZ}-Compression in Sublinear Time and Space},'' \emph{Algorithmica},
  vol.~71, no.~4, pp. 969--988, 2015.

\bibitem[Nakashima et~al.(2015)Nakashima, I, Inenaga, Bannai, and
  Takeda]{Nakashima2015CLT}
Y.~Nakashima, T.~I, S.~Inenaga, H.~Bannai, and M.~Takeda, ``{Constructing
  {LZ78} Tries and Position Heaps in Linear Time for Large Alphabets},''
  \emph{Inform.\ Process.\ Lett.}, vol. 115, no.~9, pp. 655 -- 659, 2015.

\bibitem[Fischer et~al.(2015)Fischer, I, and K{\"{o}}ppl]{lzciss}
J.~Fischer, T.~I, and D.~K{\"{o}}ppl, ``{Lempel-Ziv Computation in Small Space
  (LZ-CISS)},'' in \emph{CPM}, 2015, pp. 172--184.

\bibitem[Munro(1996)]{munro96tables}
J.~I. Munro, ``Tables,'' in \emph{Proc.\ FSTTCS}, ser. LNCS, vol. 1180.\hskip
  1em plus 0.5em minus 0.4em\relax Springer, 1996, pp. 37--42.

\bibitem[Raman et~al.(2007)Raman, Raman, and Satti]{rrrBV}
R.~Raman, V.~Raman, and S.~R. Satti, ``{Succinct Indexable Dictionaries with
  Applications to Encoding K-ary Trees, Prefix Sums and Multisets},'' \emph{ACM
  Trans. Algorithms}, vol.~3, no.~4, 2007.

\bibitem[Franceschini et~al.(2007)Franceschini, Muthukrishnan, and
  P\v{a}tra\c{s}cu]{RadixSortNoExtra}
G.~Franceschini, S.~Muthukrishnan, and M.~P\v{a}tra\c{s}cu, ``{Radix Sorting
  with No Extra Space},'' in \emph{{ESA}}.\hskip 1em plus 0.5em minus
  0.4em\relax Springer, 2007, vol. 4698, pp. 194--205.

\bibitem[Grossi and Vitter(2005)]{phiarray}
R.~Grossi and J.~S. Vitter, ``{Compressed Suffix Arrays and Suffix Trees with
  Applications to Text Indexing and String Matching},'' \emph{SIAM Journal on
  Computing}, vol.~35, no.~2, pp. 378--407, 2005.

\bibitem[Jacobson(1989)]{jacobson89space}
G.~J. Jacobson, ``Space-efficient static trees and graphs,'' in \emph{Proc.\
  FOCS}.\hskip 1em plus 0.5em minus 0.4em\relax IEEE Computer Society, 1989,
  pp. 549--554.

\bibitem[Sadakane(2007)]{cstsada}
K.~Sadakane, ``{Compressed Suffix Trees with Full Functionality},''
  \emph{{Theory of Computing Systems}}, vol.~41, no.~4, pp. 589--607, 2007.

\bibitem[Navarro and Sadakane(2014)]{statdynminmax}
G.~Navarro and K.~Sadakane, ``{Fully Functional Static and Dynamic Succinct
  Trees},'' \emph{ACM Trans. Algorithms}, vol.~10, no.~3, pp. 16:1--16:39,
  2014.

\bibitem[Sadakane and Grossi(2006)]{sadakane06squeezing}
K.~Sadakane and R.~Grossi, ``Squeezing succinct data structures into entropy
  bounds,'' in \emph{Proc.\ SODA}.\hskip 1em plus 0.5em minus 0.4em\relax
  ACM/SIAM, 2006, pp. 1230--1239.

\end{thebibliography}

\JO[\newcommand{\myDoc}{document}\end{\myDoc}]{}
\newpage
\appendix
\Section{Appendix}

\SubSection{Overview of used data structures}

While describing both factorization algorithms, we used several data structures, 
among others bit vectors, some with rank or select-support, to achieve the small space bounds.
We give here an overview (see also~\Cref{tableDSOverview}).
We denote bit vectors with~$\bv{\alpha}$ for some letter $\alpha$. %

We use for both factorizations
\begin{itemize}
	\item \bv{W} marking all witness nodes,
	\item the array~$W$ mapping witness ids to 
		\begin{itemize}
			\item (LZ77) text positions, or
			\item (LZ78) factor indices.
		\end{itemize}
\end{itemize}

In LZ77 we use
\begin{itemize}
	\item \bv{V} marking visited nodes, and
\end{itemize}

In LZ78 we use
\begin{itemize}
	\item \bv{C} marking corresponding leaves (is used as a counter),
	\item \bv{J} marking jump nodes 
	\item \bv{V} marking suffix tree nodes represented in the LZ trie (their ingoing edges got fully explored),
	\item the array~$J$ storing $n/\lg n$ numbers binary,
	\item \bv{E} marking explicit LZ trie nodes, and
	\item \bv{U} as a copy of \bv{V}.
\end{itemize}

\begin{table}[t]
	\centerline{%
		\begin{tabular}{lllllllccc}
			Common\\
			\toprule
			Name       & bits                          &  &  &         & $\rank$ & $\select$ & compress
			\\\midrule
			\bv{W}     & ${\Entr{n,z}}n+{\oh{n}}$      &  &  &         & \maru{} & \Maru{}   & \maru{}
			\\\midrule
			\ST        & $4n + {\oh{n}}$               &  &  & \Maru{} & \Maru{} & \Maru{}   &
			\\\midrule
			$\psi$     & {\Oh{n \lg \sigma}}           &  &  & \Maru{} & \Maru{} & \Maru{}   &
			\\\bottomrule
			\\
			LZ77\\
			\toprule
			Name       & bits                  & \ref{it77pass1} & \ref{it77pass2} && $\rank$ & $\select$ & compress
			\\\midrule
			\bv{V}     &                       & \maru{}         & \maru{}         && \Maru{} & \Maru{}   & \Maru{} \\\midrule
			$W$        & $z\lg n$              & \maru{}         & \maru{}         && \Maru{} & \Maru{}   & \Maru{} 
			\\\bottomrule
			\\
			LZ78
			\\\toprule
			Name       & bits         & \ref{it77pass1} & \ref{it77pass2} & \ref{it78pass3} & $\rank$ & $\select$
			\\\midrule
			\bv{C}     &              & \maru{}         & \maru{}         & \maru{}         & \Maru{} & \Maru{} \\\midrule
			\bv{J}     & $n+{\oh{n}}$ & \maru{}         & \maru{}         & \maru{}         & \maru{} & \maru{} \\\midrule
			\bv{V}     &              & \maru{}         & \maru{}         & \maru{}         & \Maru{} & \Maru{} \\\midrule
			\bv{E}     & $z+{\oh{z}}$ & \Maru{}         & \Maru{}         & \maru{}         & \Maru{} & \maru{} \\\midrule
			\bv{U}     &              & \Maru{}         & \Maru{}         & \maru{}         & \Maru{} & \Maru{} \\\midrule
			$J$        &              & \maru{}         & \maru{}         & \maru{}         & \Maru{} & \Maru{} \\\midrule
			$W$        & $z\lg z$     & \maru{}         & \maru{}         & \maru{}         & \Maru{} & \Maru{}            & \Maru{}  %
			\\\bottomrule
		\end{tabular}
	}%
	\caption{Additional data structures used while computing a LZ77/78 factorization.
		The letters written in brackets represent a pass (e.g.,~\ref{it78pass1} refers to \Cref{it78pass1}).
		The number of bits is omitted if it is exactly $n$. Circles symbolize that the data structure is used during a pass, or that it is used with 
		a $\rank$ or $\select$ structure, or that the data structure is compressible.
	}
\label{tableDSOverview}
\end{table}

\end{document}